\documentclass[preprint,aps,psfig,nofootinbib,showpacs]{revtex4}
\usepackage{bm}
\usepackage{epsfig}
\bibstyle{apsrev}
\def\gev{{\rm \,Ge\kern-0.125em V}}

\def\half{{1\over2}}

\def\Fac{{\rm Fac}}

\def\dslash{\mathbin{\partial\mkern-10mu\big/}}
\def\dis{}
\begin{document}

\begin{flushright}
\tt hep-ph/0404284v2
\end{flushright} 
\title{The Kinetic Equation for Electroweak Baryogenesis}
\author{Jitesh Bhatt$^*$ and Raghavan Rangarajan$^\dagger$
}
\affiliation{Theoretical Physics Division, Physical Research Laboratory\\
Navrangpura, Ahmedabad 380 009, India}
\date{\today}
\begin{abstract}
We derive the kinetic equation for fermions and antifermions interacting
with a planar Higgs bubble wall during the electroweak phase transition
using the `evenisation' procedure.  Equations of motion in a relativistic
quantum theory do not mirror classical relations unless one uses evenised
operators.  We give a brief introduction to evenisation and then use
the evenised Heisenberg equations of motion 
to obtain the velocity and force for the particles in the presence of
the Higgs bubble wall.  
Keeping quantum contributions to $O(\hbar)$ in
the equations of motion we obtain the semi-classical force obtained earlier
by other techniques.

\vspace{0.7cm}
{\tt
\noindent 
$^*$ jeet@prl.ernet.in\\
$^\dagger$ raghavan@prl.ernet.in\\
}
\end{abstract}
\maketitle
\newpage
\setcounter{page}{1} 
\section{Introduction}

Electroweak baryogenesis allows the possibility that the observed 
matter-antimatter
asymmetry  of the universe may have been obtained during the electroweak
phase transition at temperatures of 
around 100 GeV, 
corresponding to an energy scale that is experimentally accessible.  Hence there
has been considerable interest in  ascertaining the details of electroweak 
baryogenesis
models, in particular, the kinetic equation or Boltzmann equation for
the transport of particles through the Higgs bubble wall in a first order
electroweak phase transition.

It is well known that the characteristics of the
Boltzmann equation, in the absence of
collisions, are the single particle trajectory equations, 
i.e. if $f_{}({\bf p},{\bf x},t)$ is a distribution 
function for a system of particles 
then $\frac{df_{}({\bf p},{\bf x},t)}{dt}=0$, 
where $\bf x$ and $\bf p$ satisfy single particle
equations of motion.  Therefore expanding the total derivative
using the chain rule 
as
\begin{equation}
\partial_tf_{}\,+\frac{d {\bf x}}{dt}\partial_{\bf x} f_{}\,
+\frac{d{\bf p}}{dt}\partial_{{\bf p}}f_{}\,=\,0 \, ,
\label{BE}
\end{equation}
the quantities
$\frac{d {\bf x}}{dt}$ and $\frac{d {\bf p}}{dt}$ can be obtained
for the system
from the single particle equations of motion (Hamilton's equations).
However, in relativistic quantum theories it is non-trivial
to obtain the single particle equations of motion from the Hamiltonian
because of interference between particle and antiparticle states causing
Zitterbewegung.  Obtaining equations of motion that can have a consistent
single particle interpretation
requires the use of `evenised' operators \cite{grei}, as we discuss below.
We apply the procedure of evenisation to calculate 
particle trajectories to $O(\hbar)$
in the presence of the Higgs bubble
wall during the electroweak phase transition (presumed to be first
order).  We then use this to obtain the kinetic equation.

In Ref. \cite{prok} the energy 
relation, $\omega_{s \pm}({\bf p},{\bf x})$, and 
a kinetic equation 
in the absence of collisions
for particles
and antiparticles 
interacting with the Higgs bubble wall are obtained
from the equation of motion for the Wigner function $G^<(x,k)$,
which is the Wigner transform of the Wightman function 
$i\left< \bar{\psi}(x')\psi(x'') \right>$.
(Also see Ref. \cite{prok2}.) 
In this approach, particle and antiparticle distribution functions
are extracted from the Wigner function using a spectral decomposition that 
separates positive and negative energy states.
The particle and antiparticle distribution functions
are then found to
satisfy a collisionless quantum corrected kinetic equation
in the presence of the external force field provided by the Higgs
wall.

The kinetic equation has also been derived in Ref. \cite{cline}.
The energy relation was obtained using an WKB ansatz for components of the
(spinor) wave function.  The single particle trajectory was obtained
by treating the energy relation as the Hamiltonian.
Earlier analyses of the kinetic equation using the WKB ansatz were developed
in Refs. \cite{jpt282,cline1}.

In the above approaches the dominant contribution to the CP violating source
that leads to a baryon asymmetry is associated with the force term in the kinetic 
equation.
Alternate formalisms for the
derivation of the kinetic equation have been provided in 
Refs. \cite{huetnelson,riotto1,riotto2,riotto3,care}.   
In Refs. \cite{huetnelson} and \cite{riotto1} a source term was included
in the diffusion equation to describe the CP violating interaction between
the fermions/antifermions and the bubble wall in the presence
of the thermal plasma.  In the former, the source
was calculated by including reflection and transmission of particles and 
antiparticles by each layer of the wall.  In Ref. \cite{riotto1} the expectation
value of the CP violating current was calculated using the Closed Time Path
formalism and its divergence was included as a source.
In Refs. \cite{riotto2,riotto3} the Closed Time Path formalism was used
to derive the kinetic equation itself from the equation of motion of the 
relevant Green's 
function.

The kinetic equation obtained below by us agrees with Ref. \cite{prok2}.
In particular we obtain the semi-classical force term, first discussed in
Ref. \cite{jptcfm}.  
Our approach provides an 
alternate and intuitive derivation of the (collisionless)
kinetic equation relevant
to electroweak baryogenesis to $O(\hbar)$.  
Furthermore, we are not aware of earlier
attempts to use evenisation to study particle trajectories at higher than
$O(\hbar^0)$.

\section{}
A first order electroweak phase transition proceeds
via the formation of Higgs bubbles.  As the bubbles expand they move through the
ambient sea of quarks, leptons and other particles.  To calculate the baryon
asymmetry created as the ambient particles interact with the expanding Higgs bubble it
is necessary to study the kinetic equation
describing the passage of the particles through the bubble wall. 
Below we
first obtain an expression
for the single particle
energy to $O(\hbar)$ in terms of the evenised position and momentum.  
We then obtain expressions for the evenised velocity and force and
substitute these in Eq. (\ref{BE}) to obtain the kinetic equation.

We first give a brief introduction to evenisation.
In relativistic quantum mechanics the eigenvalues of many operators
do not agree with their classical values.  For example, for a free particle 
the eigenvalues of
the velocity operator, 
$d{\bf\hat x}/dt=(i\hbar)^{-1}[{\bf\hat x},\hat H]=
c{\bf\hat \alpha}$,
are $\pm c$ even for a massive particle.  In addition,
one can not obtain relations analogous to classical equations
for expectation values of
quantum operators 
(i.e. Ehrenfest's
theorem), as also evidenced by the expression for the velocity operator
(quantum mechanically $d\langle{\bf\hat x}\rangle/dt=
c\langle{\bf\hat \alpha}\rangle$
while classically, $d{\bf x}/dt={\bf p}/E$).  All this is
because of interference between particle and antiparticle states leading to
Zitterbewegung \cite{grei}.
One may resolve this by considering `even' parts of
operators, and by replacing operators by their `even' counterparts in equations
of motion.  This then allows a consistent
single particle interpretation of the Dirac equation.

As an illustration, the evenised velocity operator is given by
$[d{\bf \hat x}/dt]= [\hat\alpha]=
{\bf \hat p}\hat \Lambda/\sqrt{{\bf \hat p}^2+m^2}$,
where $\hat \Lambda$ is a sign operator with eigenvalues
$\pm 1$ (and we now set $c=1$).  This mimics the classical
relation.
Analogously, we shall obtain an expression for the 
Hamiltonian (actually its square) in terms of evenised position and momentum
operators and use that to obtain an  
expression for the energy of the
particle.  We will further obtain expressions for the 
evenised velocity and 
force on the particle in the presence of the bubble wall 
and use these to obtain the kinetic equation.

Any operator $\hat{A}$ can be split into an even part $\left[\hat{A}\right]$ 
and into
an odd part  $\left\{\hat{A}\right\}$. Even and odd parts of the operators are 
defined by
using the sign operator \cite{grei},
\begin{equation}
 \hat{\Lambda}\,=\,\frac{\hat{H}}{\sqrt {\hat H^2}} \, ,
\end{equation}
where $\hat H$ is the Hamiltonian.
The eigenvalues of the sign operator are $\pm1$, corresponding to particle and
antiparticle states.  Then
\begin{equation}
{\dis \left[\hat{A}\right]=\frac{1}{2}\left({\hat{A}}+\,
\hat\Lambda{\hat{A}}\hat\Lambda\right)}
\end{equation}
\begin{equation}
{\dis \left\{\hat{A}\right\}=\frac{1}{2}\left({\hat{A}}
-\,\hat\Lambda{\hat{A}}\hat\Lambda\right)}
=\frac{1}{2} [\hat A,\hat \Lambda]\hat\Lambda
\label{oddA}\end{equation}
The even part of the product of two operators $\hat{A} $ and ${\hat{B}}$ 
can be written as
\begin{equation}
\left[\hat{A}\hat{B}\right]\,=\,\left[\hat{A}\right]\left[\hat{B}\right]\,+
\left\{\hat{A}\right\}\left\{\hat{B}\right\} 
\label{prod}\end{equation}
We now apply the above to the system under consideration.

The Lagrangian describing the
interaction of particles with the bubble wall can be modeled by
\begin{equation}
{\cal L}=i\bar\psi \dslash\psi -m_R\bar\psi \psi 
-i m_I\bar\psi\gamma_5\psi
\label{lagrangian}\end{equation}
\noindent
where $m_R$ and $m_I$ are real numbers.\footnote
{Alternate ways of writing the mass terms are
$-|m|\bar\psi e^{i\theta\gamma_5}\psi$ and
$-m\bar\psi P_R\psi - m^*\bar\psi P_L\psi$, where $m=|m|e^{i\theta}
=m_R +i m_I$, as in Refs. \cite{prok,cline}.}
The Higgs bubble will be treated as a 
background field which provides a complex spatially varying mass for the particles
in the bubble wall frame, in the limit of large bubbles when the walls can be treated
as planar.
The corresponding Dirac equation 
\begin{equation}
(i \dslash -m_R
-i m_I\gamma_5)\psi=0
\end{equation}
can be rewritten in the form
\begin{equation}
i\frac{\partial \psi}{\partial t} = \hat H \psi \, ,
\end{equation}
with $\hat{H}$ given by
\begin{equation}
{\ \hat{H}}={\bf \hat{\alpha}\cdot\,\hat{p}}+\hat{\beta}\hat m_R\,
+\,i\hat{\gamma^0}\hat{\gamma^5}
\hat m_I \, .
\end{equation}
We now use the method of evenisation 
and evenise $\hat{H}^2$.  We discuss later why we 
evenise $\hat{H}^2$ rather than $\hat{H}$.

We first write after some algebra,
\begin{equation}
\hat{H}^2\,=\,{\bf{\hat{p}}}^2\,+\,|\hat m|^2\,+{\bf{\hat{\alpha}}}\hat{\beta}\left[{\bf{\hat{p}}},\hat m_R\right]
\,+\,i{\bf{\hat{\alpha}}}\hat{\gamma}^0\hat{\gamma}^5\left[{\bf\hat{p}},\hat m_I\right]
\end{equation}
\noindent
where, $|\hat m|^2\,=\,\hat m^2_R+\hat m^2_I$.  Note that $\bf\hat\alpha$ and 
$\bf\hat p$ are contracted
in the last two terms on the right hand side.
We take the wall to be planar in the $x-y$ plane and so 
$\hat m$ is a function of $\hat z$ only.
Then the Hamiltonian can be written as
\begin{equation}
\hat{H}^2\,=\,{\bf \hat{p}}^2\,+\,|\hat m|^2\,-(\hat{\alpha^3}\hat\gamma^5)
\hat\gamma^0\hat\gamma^5(-i\hbar\partial_z \hat m_R)\,
-i(\hat{\alpha^3}\hat\gamma^5)\hat{\beta}(-i\hbar\partial_z \hat m_I) \, ,
\end{equation}
where $z\equiv x^3$ and $\partial_z\equiv\partial/\partial \hat x^3$.
Defining the spin operator ${\bf\hat{S}}$ as ${\hat{\alpha}}\hat\gamma^5$ (without
the standard $\hbar/2$ in the definition to be consistent with the notation in
Refs. \cite{prok,cline}; the spin operator defined here is 
$\bf \hat \Sigma$ with Pauli matrices along the diagonal) 
the Hamiltonian can be rewritten as
\begin{equation}
\hat{H}^2\,=\,{\bf \hat{p}}^2\,+\,|\hat m|^2\,-\hat{S^3}
\hat\gamma^0\hat\gamma^5(-i\hbar\partial_z \hat m_R)\,
-i\hat{S^3}\hat{\beta}(-i\hbar\partial_z \hat m_I) \, .
\label{Hsquared}
\end{equation}
We chose to express $\hat\alpha^3\hat\beta$ 
and $\hat\alpha^3\hat\gamma^0\hat\gamma^5$
in terms of $\hat S^3$
as this simplifies the evenisation of the corresponding terms below.

We now evenise $[\hat H^2]$ to order $\hbar$.  
The spin operator in general does not commute with the Hamiltonian.
In order to simplify the problem, as in Ref.\cite{prok}, we assume
that we are in an inertial reference frame where  $x$ and $y$ 
components of the momentum
are zero and consequently $\hat \alpha^1$ and $\hat \alpha^2$ will be absent from
the Hamiltonian.  This will allow us to set $\{S^3\}$ to 0 later.
As mentioned earlier, we wish to express $[\hat H^2]$ in terms of
$[\hat p_z]$ and $[\hat z]$ (where $p_z\equiv p^3$).  The odd part of $\hat p_z$ will be proportional
to $[\hat\Lambda,\hat p_z]$ (see Eq. (\ref{oddA}) above) and so will be of order
$\hbar$.  Therefore, using Eq. (\ref{prod}), we can approximate $[\hat p_z^2]$
by $[\hat p_z]^2$.  Similarly we approximate $[m_R^2(\hat z)+m_I^2(\hat z)]$ as
$[m_R(\hat z)]^2+[m_I(\hat z)]^2$.  Furthermore, since $\{\hat z\}$ is $O(\hbar)$,
$[m_R(\hat z)]=m_R([\hat z])+O(\hbar^2)$ as any dependence of the even operator
$[m_R(\hat z)]$ on the odd operator $\{\hat z\}$ can only appear as
$\{\hat z\}^2$.

The last two terms on the right hand side of Eq. (\ref{Hsquared}) are 
$O(\hbar)$
and therfore we will use the sign operator defined upto $O(\hbar^0)$ to
evenise these terms.
We
define the zeroth order energy as
\begin{equation}
E_0={\dis \sqrt{{{p_z}}^2\,+\,|m|^2}} \,
\end{equation}
where ${p_z}^2$ and $|m|^2$ are real numbers and are expectation values
of the corresponding operators in an eigenstate of definite energy and spin.
( $\hat{}$
designates an operator.)
\noindent
With this we define 
the sign operator upto $O\left(\hbar^0\right)$ as follows
\begin{equation}
{\dis \hat{\Lambda}_0\,=\,\frac{\hat{H}}{\sqrt{{{p_z}}^2\,+\,|m|^2 }}}
\end{equation}
Note that $( \hat{\Lambda}_0)^2= 1+O(\hbar) $ which we shall use in the
derivation of evenised operators below.

Now 
\begin{equation}
\,[\hat S^3]\,=\,[\hat{\alpha^3}\hat\gamma^5]\,=\,\hat S^3 \, ,
\end{equation}
i.e., $\{\hat S^3\}=0$ as $\hat S^3$ commutes with the Hamiltonian
(which is true only in this preferred inertial frame),
and
\begin{equation}
[\hat \gamma^0\hat \gamma^5]\,=\,-i\frac{\hat m_I}{E_0}\hat \Lambda_0\,+\,O(\hbar)
\end{equation}
\begin{equation}
[\hat\beta]\,=\,\frac{\hat m_R}{E_0}\hat\Lambda_0\,+\,O(\hbar) \, .
\end{equation}
Derivations of the above expressions are provided in the Appendix.
Thus
\begin{equation}
[\hat{H}^2]\,=\,[\hat{p_z}]^2\,+
\, \hat m_R^2 +\hat m_I^2\,
-\hbar\hat S^3
\frac{\hat m_I}{E_0}\partial_z \hat m_R \hat\Lambda_0\,
-\hbar \hat S^3 \frac{\hat m_R}{E_0}\partial_z \hat m_I \hat \Lambda_0
\end{equation}
where all mass functions are functions of $[\hat z]$.
Replacing evenised operators $[\hat z]$ and $[\hat p_z]$ by 
real numbers representing their expectation values in
states of definite energy and spin,
$|E,s\rangle$ for particles and $|-E,-s\rangle$ for antiparticles,
we deduce the corresponding expression for the energy to be
\begin{equation}
E^2\,=\,{p_z}^2+|m|^2-\frac{\hbar s}{E_0}
(m_I\partial_zm_R-m_R\partial_zm_I)\, ,
\end{equation}
where we have replaced $\hat S^3$ by its eigenvalues 
$\pm s$ and 
$\Lambda_0$ by $\pm1$ for particle/antiparticle states.  
\noindent
Writing $m_R=|m|\cos\theta $ and $m_I=|m|\sin\theta $ one can rewrite the above
equation as
\begin{equation}
E^2\,=\,{p_z}^2+|m|^2-\frac{\hbar s}{E_0}|m|^2{\theta}^\prime
\end{equation}
The energy relation for both particles and antiparticles
can then be written upto order $\hbar$ as
\begin{equation}
E\,=\,\sqrt{{p_z}^2+|m|^2}-\frac{\hbar s}{2E_0^2}|m|^2{\theta}^\prime
\label{energy}\end{equation}
This expression for the energy in terms of the position, momentum and spin
of the particles/antiparticles
is the same as the energy relation obtained in Ref. \cite{prok2}.
Note that if
we had wished to obtain the energy relation to $O(\hbar)$ by evenising
$\hat{H}$ instead of evenizing $\hat{H}^2$ we would have faced a problem as 
we would have needed
$\hat \Lambda$ to $O(\hbar)$ which itself
requires $E$ to $O(\hbar)$.  
However, 
since $\{\hat H\}=0$, $[\hat H]=[\hat H^2]^{1/2}$.
We are now able to define the sign operator upto $O(\hbar)$ as
$\hat\Lambda=\hat H/E$ and use the same to obtain the evenised velocity and
force to $O(\hbar)$ for the kinetic equation.

We now obtain the kinetic equation.
Using the chain rule for partial differentiation
the kinetic equation, in the absence of collisions, is written as
\begin{equation}
\partial_tf_{s\pm}\,+\frac{dz}{dt}\partial_z f_{s\pm}\,
+\frac{dp_z}{dt}\partial_{p_z}f_{s\pm}\,=\,0 \, ,
\label{boltzeqn1}
\end{equation}
where $f_{s\pm}$ are the particle and antiparticle distribution functions.
We will associate $dz/dt$ and $dp_z/dt$ with the expectation values of 
$[d\hat z/dt]$ and $[d\hat p_z/dt]$ in states of definite energy and spin as
before.
Implicitly we are assuming here that the form of the quantum Boltzmann equation
is the same as that of the classical Boltzmann equation and the quantum 
corrections are contained only in the coefficients of the equation, namely,
in the expression for the velocity and the force.
The evenised expressions for the velocity and force to $O(\hbar)$ are derived in
the Appendix as
\begin{eqnarray}
[{d \hat z/dt}]&=&[\,-{i\over\hbar}[\hat z,\hat H]\,]\cr
              &=&{[\hat p_z]\over E}\hat\Lambda
	      \label{vel}\end{eqnarray}
	      \begin{eqnarray}
[{d \hat p_z/ dt}]&=&[\,-{i\over\hbar}[\hat p_z,\hat H]\,]\cr
                &=&\left(-{|\hat m|^{2'}\over 2E} 
	+{\hbar \hat S^3\hat\Lambda_0 (|\hat m|^{2}\hat\theta')'\over 2 E^2 }
		\right ) \hat\Lambda\, ,
\label{force}\end{eqnarray}
where $|\hat m|$ and $\hat\theta'$ are functions of $[\hat z]$ and
$E$ is as given in Eq. (\ref{energy}).
Substituting the eigenvalues of the above
operators in the kinetic equation \footnote{
The overall minus sign in the expectation value for the velocity and force
for antiparticles can be absorbed by taking the expectation value of the
momentum in antiparticle states to be $-p_z$, 
as for systems with constant mass, where the energy eignenstates are also
momentum eigenstates.
}, 
we get
\begin{equation}
\partial_tf_{s\pm}\,+{p_z\over E}\partial_z f_{s\pm}\,
+\left(-{|m|^{2'}\over 2E} 
		+ {\hbar s (|m|^{2}\theta')'\over 2 E^2 }
		\right )\partial_{p_z}f_{s\pm}\,=\,0 \, ,
\label{boltzeqn2}
\end{equation}
We note that this agrees with the kinetic equation obtained in 
Ref. \cite{prok2}. \footnote
{In the earlier version of this paper, we had identified antiparticles of spin 
$s$ with negative energy particle solutions of spin $s$, rather than
$-s$ (see Sec. 7.1 of Ref. \cite{grei}).  
Hence the expressions for energy and force were different for particles
and antiparticles of the same spin.  Our earlier results agreed with 
Refs. \cite{prok,cline} which perhaps contain the same misinterpretation.
Our current results reflect the $P$ and $CP$ violation and $C$ conservation
properties of the Lagrangian in Eq. (\ref{lagrangian}).}

Several comments are in order here. 
If we apply
Hamilton's equations to the energy relation in Eq. (\ref{energy}),
i.e., $dz/dt=\partial E/\partial{p_z}$ and 
$dp_z/dt=-\partial E/\partial{z}$,
we get expressions
for $dz/dt$ and $d p_z/dt$ that differ from the expectation
values of the corresponding
evenised operators $[d\hat z/dt]$ and $[d \hat p_z/dt]$.
There can be two reasons for this.
Firstly, the real number
$p_z$ appearing above in Eq. (\ref{energy}) is not the canonical 
momentum.  It represents the expectation value of $[\hat p_z]$
whereas the canonical momentum would be associated with
the expectation value of $\hat p_z$.
Secondly, ignoring for now the distinction between $\hat p_z$ and 
$[\hat p_z]$, Ehrenfest's theorem for a relativistic system would imply
$d\langle\hat z\rangle /dt=\langle\partial \hat H/\partial{{\hat p_z}}\rangle$ 
and 
$d \langle \hat p_z\rangle /dt=-\langle\partial\hat H/\partial{\hat z}\rangle$,
and not 
$d\langle\hat z\rangle /dt=
\partial \langle\hat H\rangle /\partial\langle{\hat p_z}\rangle$ and 
$d \langle \hat p_z\rangle /dt=-\partial\langle\hat H\rangle/
\partial\langle{\hat z}\rangle$.

In the literature \cite{prok,cline} there has been discussion on identifying
the canonical and kinetic momentum of the particle. One may presume that since
$\hat p_z$ is identified with the $-i\hat\partial_z$ 
operator it is conjugate to 
$\hat z$ and represents the canonical momentum.  
Now our energy relation agrees with that of Ref. \cite{prok2}.
But the energy relation in Ref. \cite{prok2} is necessarily in terms of the kinetic 
momentum as 
the Wigner function invoked in Ref. \cite{prok2} is always written in
terms of the kinetic momentum of the system (see Sec. 3 of Ref. \cite{vge}).
This indicates
that our real number $p_z=\langle [\hat p_z]\rangle$ 
in the energy relation above is the kinetic momentum of Ref. \cite{prok2}.
It is not obvious to us why the evenised canonical momentum  in this problem
becomes
equivalent to the kinetic momentum.

If one wishes to define a classical Hamiltonian function
which gives the expressions for velocity and force above
by applying Hamilton's equations, 
one may define a canonical 
momentum $p_c$ by the following ansatz of Eq. (61) of Ref. \cite{prok3d}
\begin{equation}
p_c=p_z\left(1+ \frac{\hbar s\theta'}{2(p_z^2+|m|^2)^{\half}}\right)
\label{pcansatz}\end{equation}
and reexpress the energy relation in terms of $p_c$ as
\begin{eqnarray}
E\,&=&\,\sqrt{{p_z}^2(1+\frac{\hbar s\theta'}{E_0})+|m|^2}
-\frac{\hbar s\theta^\prime}{2}\\
&=&\,\sqrt{{p_c}^2+|m|^2}
-\frac{\hbar s\theta^\prime}{2}\, .
\label{energypc}\end{eqnarray}
Then
$dz/dt=\partial E/\partial{p_c}$ and, rewriting Eq. (\ref{pcansatz})
as $p_z= \Fac\, p_c$,
$dp_z/dt=(d \Fac/dt)*p_c-\Fac*\partial E/\partial{z}$.
These agree with the expressions of the evenised velocity and force
above.  Eq. (\ref{pcansatz}) agrees with the relation between kinetic
and canonical momentum in Ref. \cite{cline}, with $\alpha_{CP}$ of
Ref. \cite{cline} set to 0.

\section{Conclusion}

In conclusion, we have provided an alternate procedure for obtaining
the energy relation and the kinetic equation 
for fermions and antifermions interacting with the
Higgs bubble wall during the electroweak phase transition using the method
of evenisation.  
Our derivation of the single particle behaviour using the Heisenberg
equations of motion has allowed us to include quantum corrections
systematically to $O(\hbar)$.  This provides a
direct and intuitive way of obtaining the kinetic equation in this limit.
Our results agree with those that may be
obtained using more formal quantum field theoretic 
methods involving the Wigner function, or the WKB method.  In particular,
we rederive the semi-classical force that may be 
obtained by these other methods.

\begin{acknowledgements}
We would like to thank Tomislav Prokopec and Steffen Weinstock 
for many
clarifications about their work and
for many useful comments
regarding an earlier version of this work.
We would also like to thank B. Ananthanarayan for fruitful discussions.
R.R. would like to thank N. P. M. Nair for help with preparation of the
manuscript.
\end{acknowledgements}

\begin{appendix}
{\bf \centerline {Appendix}}

We present below the derivations of the various evenised expressions
quoted in the text.  Standard formulae that we shall use are
$\hat\beta^2=-(\hat\gamma^0\hat\gamma^5)^2=(\hat\alpha^3)^2=1$.
The symbols square brackets are used below for both commutators and evenised operators
but the difference is clear from the context.

{\bf Evenised $\hat\beta$ and $\hat\gamma^0\hat\gamma^5$ to $O(\hbar^0)$:}

\begin{eqnarray}
2[\hat\beta]&=&\hat\beta + \hat\Lambda_0 \hat\beta \hat\Lambda_0\cr
            &=&\hat\beta + {\hat\alpha^3 p^3 +\hat\beta \hat m_R 
	    +i\hat\gamma^0\hat\gamma^5 \hat m_I \over E_0}
	    \hat\beta \hat\Lambda_0\cr
	    &=&\hat\beta + \hat\beta{-\hat\alpha^3 p^3 +\hat\beta \hat m_R 
	    -i\hat\gamma^0\hat\gamma^5 \hat m_I \over E_0}
	     \hat\Lambda_0\cr
	     &=&\hat\beta + \hat\beta{
	     -\hat\Lambda_0 E_0
	     +2\hat\beta \hat m_R 
	    \over E_0}
	     \hat\Lambda_0\cr
	     &=&{2\hat m_R\over E_0}\hat\Lambda_0 \, ,
\end{eqnarray}
where we have used $\hat\Lambda_0^2= 1+O(\hbar)$ in the last 
equality.  Since the lhs is even and the sign operator on the rhs is even,
$m_R(\hat z)$ may be replaced by $[m_R(\hat z)]$, which, as explained
in the text, is equal to $m_R([\hat z])$ to $O(\hbar)$. Therefore
\begin{equation}
[\hat\beta]={ m_R([\hat z])\over E_0}\hat\Lambda_0 +O(\hbar)\, .
\end{equation}

Similarly
\begin{eqnarray}
2[\hat\gamma^0\hat\gamma^5]&=&\hat\gamma^0\hat\gamma^5 +
\hat\Lambda_0 \hat\gamma^0\hat\gamma^5 \hat\Lambda_0\cr
&=& \hat\gamma^0\hat\gamma^5 + \hat\gamma^0\hat\gamma^5
{
-\hat\Lambda_0 E_0 
+2\hat\gamma^0\hat\gamma^5 i \hat m_I 
\over E_0}\hat\Lambda_0\cr
&=&-{2i \hat m_I\over E_0}\hat\Lambda_0\, .
\end{eqnarray}
Therefore, replacing $m_I(\hat z)$ by $m_I([\hat z])$,
\begin{equation}
[\hat\gamma^0\hat\gamma^5]=-{i m_I([\hat z])\over E_0}\hat\Lambda_0+O(\hbar)\, .
\end{equation}

{\bf Evenised expressions for the velocity and the force
to $O(\hbar)$:}  

Since we wish to work till $O(\hbar)$ 
we now use the sign operator $\hat\Lambda$ defined as
\begin{equation}
\hat\Lambda={\hat H \over E}
\end{equation}
with $E$ defined to $O(\hbar)$ as in Eq. (\ref{energy}).  Now
$d\hat z/dt=-(i/\hbar)[\hat z,\hat H]=\hat \alpha^3$.  Then
\begin{eqnarray}
2[\hat\alpha^3]&=&\hat\alpha^3 + \hat\Lambda \hat\alpha^3 \hat\Lambda\cr
&=&\hat\alpha^3 + \hat\alpha^3 {
-\hat\Lambda E
+2\hat\alpha^3 \hat p^3 
\over E}
\Lambda\cr
&=&{2 \hat p^3 \over E} \hat\Lambda \, =\,{2 \hat p_z \over E} \hat\Lambda \,.
\end{eqnarray}
Since the lhs of the above equation
and the sign operator on the rhs are even we may replace
$\hat p_z$ by $[\hat p_z]$.  Therefore we get
\begin{equation}
[d\hat z/dt]={ [\hat p_z] \over E} \hat\Lambda \, .
\end{equation}

The force $d \hat p_z/dt=-(i/\hbar)[\hat p_z,\hat H]=
-\hat \beta\hat m_R' -i \hat\gamma^0\hat\gamma^5\hat m_I'$.
Then
\begin{eqnarray}
2[\hat \beta \hat m_R']&=&\hat\beta\hat m_R' 
+ \hat\Lambda \hat\beta\hat m_R' \hat\Lambda\cr
&=&\hat\beta\hat m_R' + \hat\beta\hat m_R' {
-\hat\Lambda E
+2\hat\beta \hat m_R
\over E}
\hat\Lambda  -\hat\beta\hat\alpha^3{[\hat p_z,\hat m_R']\over E }\, \hat\Lambda \cr
&=&{2 \hat m_R' \hat m_R 
+i \hbar\hat\beta\hat\alpha^3 \hat m_R^{\prime\prime} \over E} \hat\Lambda \, .
\end{eqnarray}
Again, we may replace $\hat m_R' \hat m_R$ by $[\hat m_R' \hat m_R]$ as
both sides of the equation should be even.
As $\{\hat m_R'\}\{\hat m_R\}$ is $O(\hbar^2)$, Eq. (\ref{prod}) implies that
this reduces to
$[ m_R'(\hat z)][ m_R(\hat z)]$.  This 
may further be rewritten as
$ m_R'([\hat z])m_R([\hat z])$, since as discussed earlier
in the text, the dependence on $\{z\}$ of each term in the product
will be at $O(\hbar^2)$.  For the second term in the numerator we 
replace $\hat\beta\hat\alpha^3 \hat m_R^{\prime\prime}$ by
$[\hat\beta\hat\alpha^3 \hat m_R^{\prime\prime}]$.  Since this term is
already of $O(\hbar)$ from the commutation employed in its derivation,
we can then write the product as 
$[\hat\beta\hat\alpha^3][m_R^{\prime\prime}(\hat z)]=
[\hat\beta\hat\alpha^3] m_R^{\prime\prime}([\hat z])$.  Now $\hat\beta\hat\alpha^3=\hat S^3
\hat\gamma^0\hat\gamma^5$ and since $\{\hat S^3\}=0$, $[\hat\beta\hat\alpha^3]=
\hat S^3 [\hat\gamma^0\hat\gamma^5]=
-i \hat S^3  {m_I([\hat z])\over E_0} \hat\Lambda_0$.  Therefore
\begin{equation}
[\hat \beta \hat m_R']=\left( {m_R^{2\prime}([\hat z])\over 2E }
+ {\hbar\hat S^3 \hat\Lambda_0 m_I([\hat z]) m_R^{\prime\prime}([\hat z])\over 2E^2}\right)\hat\Lambda \, .
\end{equation}

For the second term in $d\hat p_z/dt$,
\begin{eqnarray}
2[\hat \gamma^0\hat\gamma^5 \hat m_I']&=& \hat \gamma^0\hat\gamma^5 \hat m_I'
+ \hat\Lambda \hat \gamma^0\hat\gamma^5 \hat m_I'\hat\Lambda\cr
&=& \hat \gamma^0\hat\gamma^5 \hat m_I' + 
\hat \gamma^0\hat\gamma^5 \hat m_I'  {
-\hat\Lambda E
+2 i \hat\gamma^0\hat\gamma^5 \hat m_I
\over E}\hat
\Lambda  -\hat\gamma^0\hat\gamma^5\hat\alpha^3{[\hat p_z,\hat m_I']\over E} \, \Lambda \cr
&=&({-2 i\hat m_I' \hat m_I 
+i \hbar\hat\gamma^0\hat\gamma^5\hat\alpha^3 \hat m_I^{\prime\prime} \over E} \hat\Lambda \, .
\end{eqnarray}
$\hat\gamma^0\hat\gamma^5\hat\alpha^3=\hat S^3\hat\beta$.
Invoking arguments as above and substituting for $[\hat\beta]$ we get
\begin{equation}
[i \hat \gamma^0\hat\gamma^5 \hat m_I']=\left( {m_I^{2\prime}([\hat z])\over 2E} 
- {\hbar\hat S^3 \hat\Lambda_0 m_R([\hat z]) m_I^{\prime\prime}([\hat z])\over 2E^2}\right)\hat\Lambda \, .
\end{equation}

Thus
\begin{equation}
[{d\hat p_z/dt}]=\left(-{|\hat m|^{2'}\over 2E} 
	+ {\hbar\hat S^3 \hat\Lambda_0(|\hat m|^{2}\hat\theta')'\over 2 E^2} 
		\right ) \hat\Lambda\, ,
\end{equation}
where we have used 
$\hat m_R \hat m_I^{\prime\prime}-\hat m_I \hat m_R^{\prime\prime}=
(|\hat m|^{2}\hat\theta')'$.
$|\hat m|$ and $\hat\theta'$ above are functions of $[\hat z]$.

\end{appendix}



\begin{thebibliography}{8} 

\bibitem{grei} W. Greiner, {\it Relativistic Quantum Mechanics}, 
Springer-Verlag, Berlin,
Third Edition (2000) (Ch. 2).

\bibitem{prok} K. Kainulainen, T. Prokopec, M. G. Schmidt and
S. Weinstock, JHEP 0106 (2001) 31
[arXiv:hep-ph/0105295].  

\bibitem{prok2} T. Prokopec, M. G. Schmidt and
S. Weinstock, Ann. Phys. 314 (2004) 208 [arXiv:hep-ph/0312110].

\bibitem{cline} J. M. Cline, M. Joyce and K. Kainulainen, JHEP 0007 (2000) 018
[arXiv:hep-ph/0006119]; erratum, arXiv:hep-ph/0110031.

\bibitem{jpt282} M. Joyce, T. Prokopec and N. Turok, Phys. Rev. D {\bf 53}
(1996) 2958 [arXiv:hep-ph/9410282].


\bibitem{cline1} J. M. Cline, M. Joyce and K. Kainulainen, Phys. Lett. 
{\bf B417} (1998) 79 [arXiv:hep-ph/9708393]; erratum, Phys. Lett. 
{\bf B448} (1999) 321.

\bibitem{huetnelson} P. Huet and A. E. Nelson, Phys. Rev. D {\bf 53}
(1996) 4578 [arXiv:hep-ph/9506477].

\bibitem{riotto1} A. Riotto, Phys. Rev. D {\bf 53}
(1996) 5834 [arXiv:hep-ph/9510271].

\bibitem{riotto2} A. Riotto, Nucl. Phys. {\bf B518}
(1998) 339 [arXiv:hep-ph/9712221].

\bibitem{riotto3} A. Riotto, Phys. Rev. D {\bf 58}
(1998) 095009 [arXiv:hep-ph/9803357].

\bibitem{care} M. Carena, J.M. Moreno, M. Quiros, M. Seco and C.E.M. 
Wagner, Nucl. Phys. {\bf B599} (2001) 158 [arXiv:hep-ph/0011055].

\bibitem{jptcfm} M. Joyce, T. Prokopec and N. Turok, Phys. Rev. Lett. {\bf 75}
(1995) 1695 [arXiv:hep-ph/9408339].

\bibitem{vge} D. Vasak, M. Gyulassy and H.-T. Elze, Annals of Physics 173 (1987) 462.

\bibitem{prok3d} K. Kainulainen, T. Prokopec, M. G. Schmidt and
S. Weinstock, Phys. Rev. D {\bf 66} (2002) 043502
[arXiv:hep-ph/0202177].  



\end{thebibliography}
\end{document}